\documentclass[12pt,preprint]{aastex}
\usepackage{amsmath,longtable,natbib}
\usepackage{xcolor}
\usepackage{lineno}
\usepackage{float}
\usepackage[caption = false]{subfig}
\usepackage{hyperref}
\usepackage{graphicx}
\shorttitle{In-field phasing at the upgraded GMRT}
\shortauthors{Kudale et al.}
\usepackage{subfig}
\usepackage[normalem]{ulem}
\begin{document}

\title{In-field phasing at the upgraded GMRT}
\author{
Sanjay Kudale\altaffilmark{1},
Jayanta Roy\altaffilmark{1},
Jayaram N. Chengalur\altaffilmark{1,2},\\
Shyam Sharma\altaffilmark{1},
Sangita Kumari\altaffilmark{1}
}
\altaffiltext{1}{National Centre for Radio Astrophysics, Pune 411 007, India}
\altaffiltext{2}{Tata Institute of Fundamental Research, Mumbai 400 005, India}
\affil{}
%===================================================================
\begin {abstract}
In time-domain radio astronomy with arrays, voltages from individual antennas are added together with proper delay and fringe correction to form the beam in real-time. In order to achieve the correct phased addition of antenna voltages one has to also correct for the ionospheric and instrumental gains. Conventionally this is done using observations of a calibrator source located near to the target field. This scheme is sub-optimal since it does not correct for the variation of the gains with time and position in the sky. Further, since the ionospheric phase variation is typically most rapid at the longest baselines, the most distant antennas are often excluded while forming the beam. We present here a different methodology ("in-field phasing"), in which the gains are obtained in real-time using a model of the intensity distribution in the target field, which overcomes all of these drawbacks. We present observations with the upgraded Giant Metrewave Radio Telescope (uGMRT) which demonstrates that in-field phasing does lead to a significant improvement in sensitivity. We also show, using observations of the millisecond pulsar J1120$-$3618 that this in turn leads to a significant improvement of measurements of the Dispersion Measure and Time of Arrival. Finally, we present test observations of the GMRT discovered eclipsing black widow pulsar J1544+4937 showing that in-field phasing leads to improvement in the measurement of the cut-off frequency of the eclipse. 
%----------------------------------------------------\\
\end {abstract}

\vskip 1.6cm
%########################   INTRODUCTION  ####################################
\par
\section{Introduction}
\label{sec:intro}
% --------------------------------------------------------------------
In radio astronomy calibration to correct for instrumental and ionospheric perturbations to the amplitude and phase of the received astrophysical signal is essential. In particular, at low radio frequencies, it is important to calibrate the ionospheric phases quite frequently. This is typically done by interleaving observations of a nearby (i.e. to the target field) calibrator source with the target field observation. In post-processing, these calibrator observations are used to interpolate the phase corrections which are then applied to the data. For time domain astronomy typically a high-time resolution data stream is provided by combining the voltages of the antennas in the array in real time to produce a "phased array" or "tied array" beam. Conventionally, for such observations, a nearby calibrator is observed just prior to the target observations in order to "phase up" the array. This scheme suffers from a number of drawbacks. Firstly, it does not allow for correction of the variation of the phases with time and position in the sky. Secondly, since the ionospheric phase varies most rapidly on the longest baselines, typically the most distant antennas are excluded while forming the beam. All of this leads to a significant loss in sensitivity. Finally, this external calibration needs to be done quite frequently, typically once in 30-40 minutes. This not only leads to increased observation overheads (about 15-20\%), but also makes it not possible to have long uninterrupted observations of the target source.  This can be a limiting factor for some studies, e.g. observations of binary and eclipsing pulsars for scanning full orbital phase (typically few hours long), repeating Fast Radio Bursts (FRBs), faint pulsars which need few hours of phase coherent folding for detection, intermittent pulsars etc.

For imaging observations, it has long been the practice to iteratively improve the dynamic range of the image by estimating antenna based gains from the target observations themselves, a process known as "self-calibration" (\cite{Cornwell1999}, 
\cite{Thompson2017}). Here we present a real time implementation of this at the upgraded GMRT (uGMRT, \citep{Gupta2017}) where a pre-existing model of the target field is used to determine and apply the phases while forming the phased array beam. We also present observations comparing in-field calibration with conventional calibration, which shows that in-field calibration does indeed lead to significant improvement in the sensitivity, which in turn leads to improvements in measurements of the Dispersion Measure (DM), Time of Arrival (TOA), as well as the cut-off frequency of the eclipse for eclipsing pulsars.

 The rest of the paper is organized as follows. We describe the methodology in section \ref{sec:Methodology}. The Observation and analysis details 
 are presented in section \ref{sec:Observation-and-analysis}. In section \ref{sec:results} we present observations taken to validate the in-field phasing implementation, as well as to compare it with conventional phasing.
 Observations of two GMRT discovered millisecond pulsars (MSPs) are presented in section \ref{sec:Application2MSP}, allowing a translation of the improvement in the sensitivity into increased accuracy of parameters of interest such as the DM, TOA, eclipse cut-off frequency etc.  We summarise the results in section \ref{sec:Discussion}.
% --------------------------------------------------------------------
%##############################################################################
\section{Methodology}
\label{sec:Methodology}
%------------------------------------------------
\par
Radio telescope arrays measure {\it visibilities}, i.e. the pairwise cross-correlation between the signals from all antennas.  In the absence of baseline based errors and non-isoplanatic effects, the observed visibilities $V_{ij}$ can be written as in equation \ref{eq:visibility}, where $\widehat{V_{ij}}$ are true visibilities, $g_i$ are antenna based complex gains and $\eta_{ij}$ is the noise. Note that the gain includes both instrumental as well as ionospheric contributions.
  \begin{equation}
  \label{eq:visibility} 
   V_{ij} (t) = g_{i}(t)g^{*}_{j}(t) \widehat{V}_{ij}(t) + \eta_{ij}(t)
 \end {equation}

If the true visibilities $\widehat{V}_{ij}$ are known, then this equation can be used to estimate the antenna based gains by minimizing
\begin{equation}
\label{eq:minimize}
L =\sum_{i=1}^{N} \sum_{j=i+1}^{N} w_{ij}|V_{ij} - g_{i}g_{j}^{*}\widehat {V}_{ij} | ^ 2
\end{equation}

where $w_{ij}$ are suitable weights. In the most straight forward case, observations of a known isolated point source (i.e. for which $\widehat{V}_{ij}$ is a constant) can be used for solving for the antenna gains. If one assumes that the gains vary only slowly with time and position in the sky, these solutions can also be used to calibrate the target visibilities. However, as is well known, for an N element array the  number of complex unknowns in Eqn.~\ref{eq:minimize} is much smaller than the number of measurements ($\sim N^2$), and schemes to iteratively solve for both the true visibilities as well as the unknown complex gains (i.e. self-calibration,   \cite{Cornwell1999}) converge rapidly. As such Eqn.~\ref{eq:minimize} can be used to determine the gain for quite complex fields, even if one starts with a relatively crude initial estimate of the gains. In situations where the true visibilities are already known (for e.g. from an earlier observation), Eqn.~\ref{eq:minimize} can be used to directly estimate the antenna based gains without the need for observations of a standard calibrator source. In particular, for phased array observations of a field in which the true visibilities are known Eqn.~\ref{eq:minimize} can be used  not just to directly phase the array (thus eliminating the error introduced by the ionospheric phase difference between the target and calibrator source directions), but also to correct for any time variability since the gains can be updated in quasi real time.

\cite{Kudale2017} had presented an implementation of this scheme for the legacy GMRT correlator, i.e. the GMRT Software Back-end (GSB, \cite{Jroy2010}). The GSB had a bandwidth of  
33 MHz band-width divided into a maximum of  512 channels. The implementation used the
\textit{flagcal} package (\cite{Prasad2012}, \cite{Chengalur2013}) to flag the data and determine the antenna based gains for individual channels by using direct Fourier transform (DFT). The upgrade of the GMRT (\cite{Gupta2017} replaced the GSB with the GMRT Wide-Band backend (GWB, \cite{Reddy2017}), which allows for up to 16384 spectral channels with a total bandwidth of up to 400 MHz.  The DFT implementation in \textit {flagcal} is sub-optimal for real-time applications, and hence an FFT-based solver was added to the package.  All the observations presented here use this updated solver.
\par
For wide fields of view, accurate calculation of the model visibilities requires one to account for non-coplanarity of the baselines (the so called "w" correction, see e.g. \cite{Cornwell2008}.  This is computationally expensive, and is not part of the current implementation. To estimate the magnitude of the error introduced, we tested out the solver on simulated 
fields with a single strong point source ($\sim$1 Jy) situated at a range of distances from the phase centre of the field.  We find that if  the source lies within $\sim$20' from phase center at band-4 (550$-$750 MHz) and within $\sim$30' from phase center at band-3 (300$-$500 MHz), the rms phase errors are less than $\sim5^{\circ}$, which is small compared to the noise. All of the observations presented here are for fields where there are no dominant sources lying beyond the above limits.  The 'w' correction would be important for sources with dominant sources at the edge of the field of view. Correction for the 'w' term is more computationally expensive and is deferred to a later implementation.
%-----------------------------------------------------------
\section{Observation and analysis}
\label{sec:Observation-and-analysis}
A number of test observations were carried out to compare the performance of in-field phasing with that of conventional phasing.  Specifically we observed the normal pulsar B0740$-$28 in the GMRT Band 4(550$-$750 MHz) and two millisecond pulsars J1120$-$3618  and J1544+4937 in the GMRT Band 3 (300$-$500 MHz), all with 200 MHz bandwidth.  J1120$-$3618 is one of the fainter MSP discovered with GMRT (\cite{Bhattacharyya22} while  MSP  J1544+4937 is an eclipsing binary pulsar (\cite{Bhattacharyya13}).
Eclipsing binaries are a special class of fast-spinning MSPs in compact binary systems, where the companions are ablated away by
energetic pulsar winds with ablated material from the companion overflowing
companion's Roche lobe (e.g. \cite{Roberts2012}). 
The intra-binary material can obscure the pulsar's emission for part of its orbit, resulting in eclipses. A black widow is a subclass of eclipsing binaries characterised by a very low companion mass ($M_c$), i.e. $M_c$ $<$ 0.1 $M_\odot$.

In order to facilitate the comparison, the observations were carried out in two modes: conventional phasing using an external calibrator source and in-field phasing.  For the conventional phasing  we observed the calibrator 0837$-$198 for B0740$-$28, 1154$-$35 for J1120$-$3618, and 1459+716 for J1544+4937. The conventional phasing was done only once, at the start of the observation. For the in-field phasing we used a sky-model for these fields obtained from earlier observations. The model visibilities obtained from this sky model, along with the real-time visibilities measured by the GWB were used to solve for per antenna phases in real-time. These phase corrections were applied to the output of FFT inside the GWB. The software allows for the phases to be updated without needing a break in the observations. 

The GWB allows for the formation of up to 4 beams along with the visibility data. For our observations we typically used three beams. The antennas whose data is to be combined can be chosen independently for each beam. The GMRT has a hybrid configuration with about half the antennas in a central compact configuration and the remaining in 3 roughly "Y" shaped arms. The first beam ("beam1") typically used antennas from the compact array and the 1st arm antenna, while the third beam ("beam3") used all the antennas, i.e. including the extreme arm antennas. The second beam ("beam2") was intermediate between these two, typically including up to the 3rd arm antenna. The actual antennas used to form each beam varied from run to run, depending on the antennas that were available for use at the given time. But overall "beam1" was restricted to relatively short baselines, "beam3" included the longest available baselines, while "beam2" is intermediate between these two. Note also that all the antennas used for producing "beam1" were also included for producing "beam2". Similarly all the antennas used for producing "beam2" are also included for producing "beam3".  Finally we note in passing that since tied arrays are formed by adding the phased voltages and squaring their output also contains a contribution from the auto-correlations of the individual antennas (i.e. the "incoherent array" output), which is independent of the phase errors (see e.g. \cite{Roy2018}).

The results presented next for the beam data were obtained via analysis using \textit{SIGPROC} (\cite{Lorimer2011}) and 
\textit{PRESTO} (\cite{Ransom2002}) for folding and \textit{TEMPO2}
(\cite{Hobbs2006}) for timing analysis. 
\par
\section{Results}
\label{sec:results}
The first set of results we obtained were from observations aimed at testing the in-field observing mode and optimising the observation strategy. We examine the variation of the SNR (signal-to-noise ratio) with time and finalize a strategy which keeps the SNR constant, as expected for in-field phasing. In the second set of observations, we use the updated observation strategy to demonstrate the scope of in-field phasing to enhance the scientific output of the GMRT for time-domain science.
%---------------------------------------------------------------
\subsection{Testing the in-field phasing scheme}
\label{subsec:de-phasing_phasing}
\begin{figure}[ht!]
\centering
%\subfloat[]{\includegraphics[width=0.5\textwidth]{SNR_de-phasing_B0740-28_28May2022_psf-flux.pdf}} 
\subfloat[]{\includegraphics[width=0.5\textwidth]{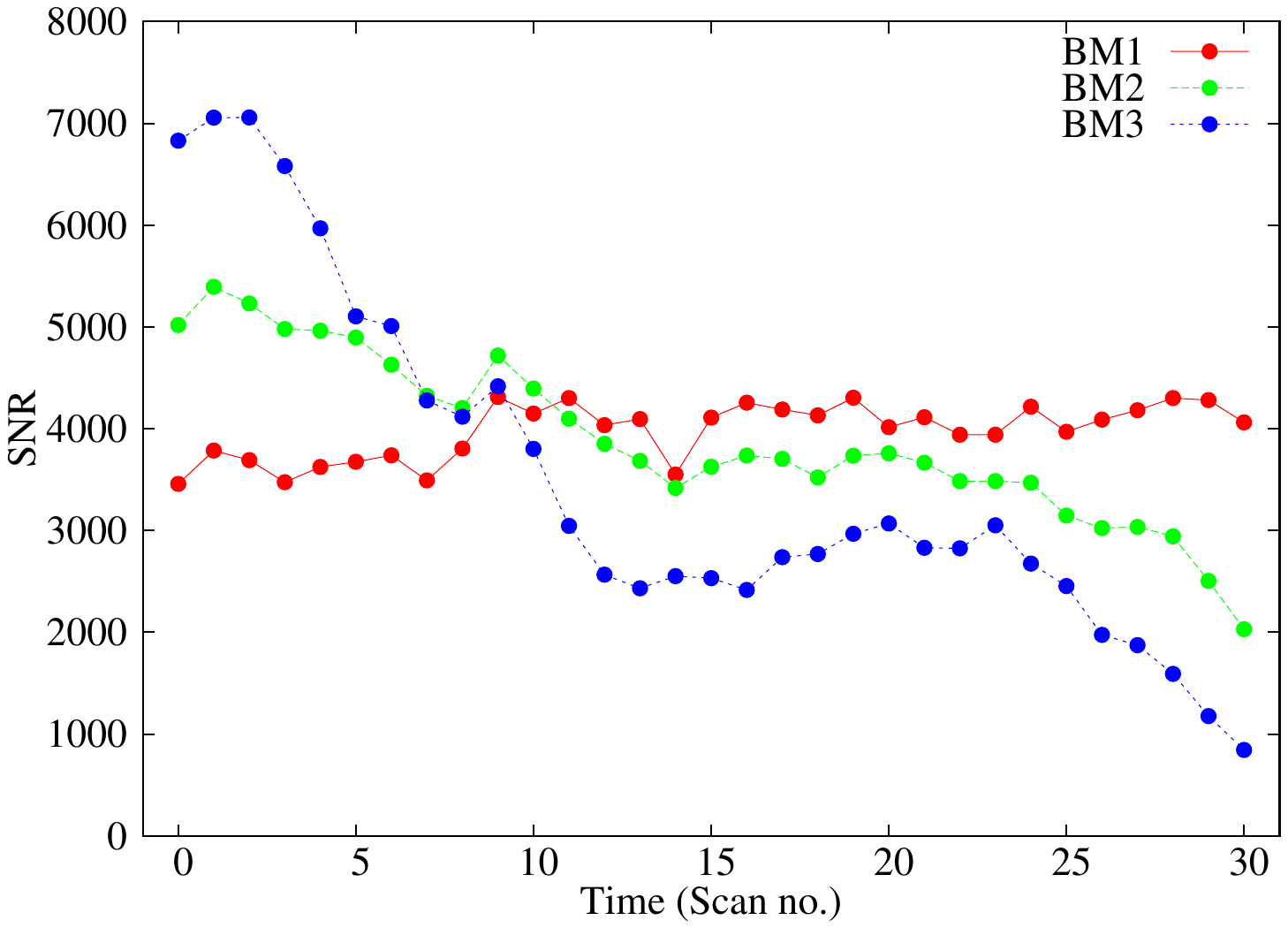}}
%\subfloat[]{\includegraphics[width=0.5\textwidth]{SNR_de-phasing_B0740-28_27Dec2023_psf_flux1.pdf}} 
\subfloat[]{\includegraphics[width=0.5\textwidth]{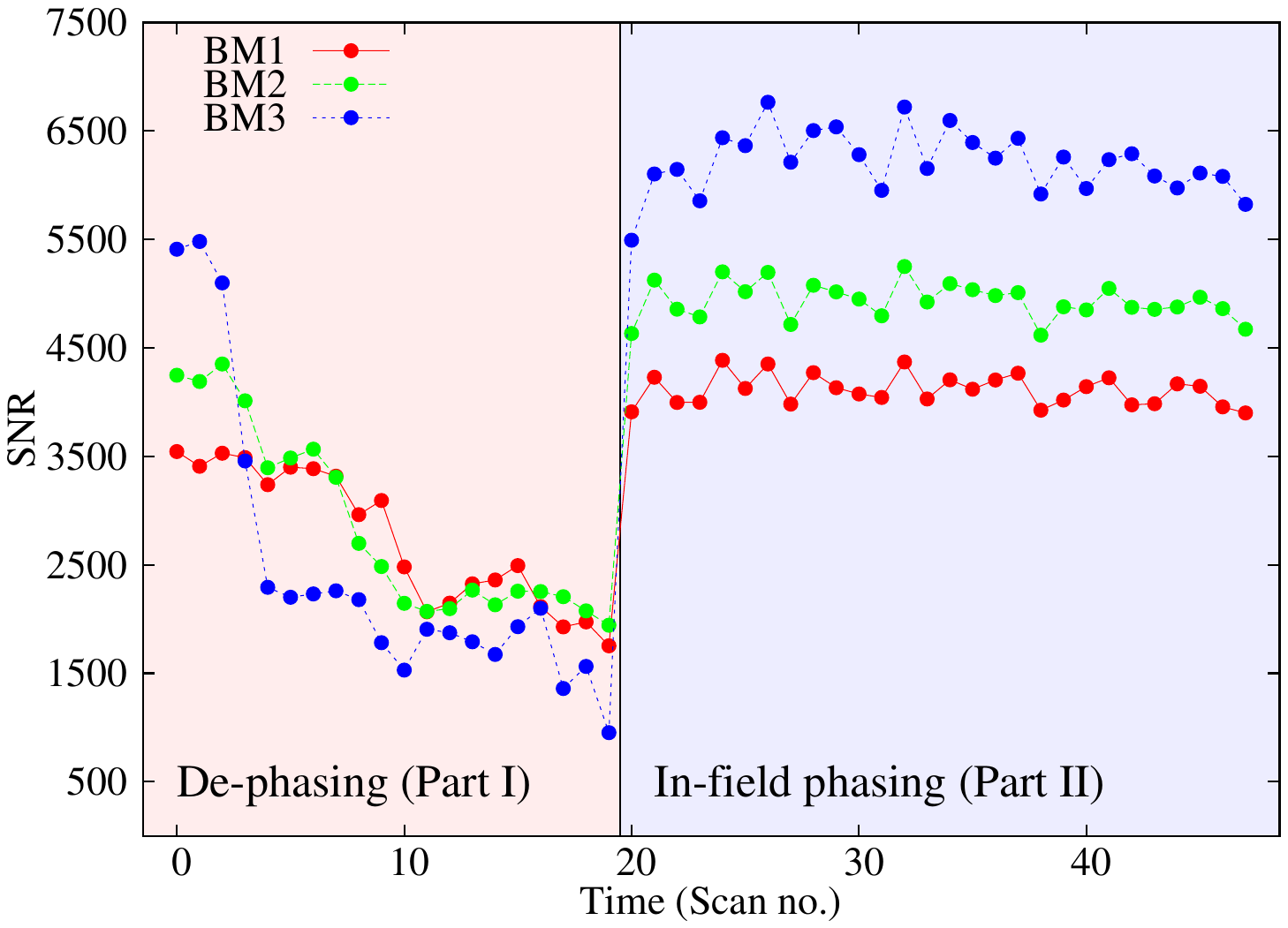}}
\caption[SNR drop due to de-phasing for different array sizes]
{Part (a) : The drop in  SNR for the different arrays as they de-phase with time. Phasing has been done only once at the start of the observation. The SNR is computed for scans $\sim$5 minute duration.
Part (b) : 
[Part I light red shaded region] The drop of SNR with time for different arrays as they de-phase with  time. Phasing has been done only once at the start of the observations.  Note that the SNR drop for this set of observations is much more rapid than for the observations shown in Part (a), showing the time variable characteristics of the de-phasing. 
[part II (light blue shaded region)]  shows the SNR when in-field phasing carried out at regular intervals. The SNR is computed for scans of $\sim 4$ minute duration.
In all panels larger array beam (beam3) is indicated by blue curve, moderate array size beam (beam2) is shown by green curve and smaller size array (beam1) is shown with red curve.  
}
\label{fig:dephasing-phasing}
\end{figure}

For probing the phasing effectiveness with in-field phasing we carried out observations on pulsar B0740$-$28 while simultaneously recording data from three beams (i.e. beam1, beam2 and beam3) with an increasing number of antennas, as described above.  At the start of the observation the array was phased using in-field phasing and for the remaining part, the pulsar data were recorded with scans of 5 minutes each for a duration of $\sim$2.5 hours without any re-phasing. The pulsar SNR was computed for each scan, and the variation of SNR with scan number is shown in Fig.~\ref{fig:dephasing-phasing}. As can be seen, beam1 (containing only short baselines) shows no drop in SNR 
%see fig \ref{fig:SNR_de-phasing_B0740-28_28May2022_psf-flux} top panel, red curve. 
(see the red curve in Part (a) of Fig. \ref{fig:dephasing-phasing}). In contrast,
beam2 (which includes antennas at intermediate distances) shows an  SNR drop 
up to $\sim$41\% of its starting SNR (see the green curve  in Part (a) of Fig. \ref{fig:dephasing-phasing}), while beam3, which includes the longest available baselines, showed an SNR drop to $\sim$12\%
%$\sim$12\% (\textbf{Give factor here rather than \%})
of its starting SNR. (see the blue curve in Part (a) of Fig.  \ref{fig:dephasing-phasing}). The figure also shows the fastest drop in SNR happens for beam3, in line with the expectation that phase variations are faster on the longer baselines. In passing we note that changes in ionospheric conditions over the array would lead to both a shift in the apparent position of the target pulsar (arising from any systematic gradients in the density) as well as a loss in sensitivity due to dephasing of the array (see e.g. \cite{Kudale2017}). Both of these effects will lead to an increasing loss of sensitivity for arrays that have a longer baselines.
%---------------------------------------------------------------
%----------------------------------------------------------------
\par
\label{sec:Results}
%-----------------------------------------------------------
\par
Figure~\ref{fig:dephasing-phasing}(b) shows another set of observations of the pulsar B0740$-$28. For the observations shown in the Part I (light red shaded region) the array was phased (using in-field phasing) only once at the start of the $\sim$1.3 hour observation.  On the other hand, for the observations shown in the Part II (light blue shaded region) the array phases were updated every 4 minutes using the data acquired during the earlier $\sim$4 minutes scan. As before the data was recorded for 3 arrays, and as can be seen without in-field phasing the SNR degrades rapidly, particularly for beam3, the array with the longest baselines. On the other hand, with in-field phasing the SNR is approximately constant, as expected. It is interesting to note that the variation of SNR with time in the case where no continuous phase correction is done is quite different for the two different observing runs separated by $\sim$1.5 years shown in Figure~\ref{fig:dephasing-phasing}.  In the observations shown in Figure~\ref{fig:dephasing-phasing}(a) the SNR even for the longest array reduces gradually over the $\sim$150 minutes observation span, whereas for the observations shown in Panel (b) of the same figure the SNR drops very sharply in the first $\sim$20 minutes. At low frequencies, where the ionosphere can sometimes be fairly dynamic, these observed sharp drops in the SNR are indicative of the limitations of conventional phasing. 
%---------------------------------------------------------------
\par
%----------------------------------------------------------
\begin{figure}[!ht]
\centering
    \includegraphics[width=3.6in,angle=0] {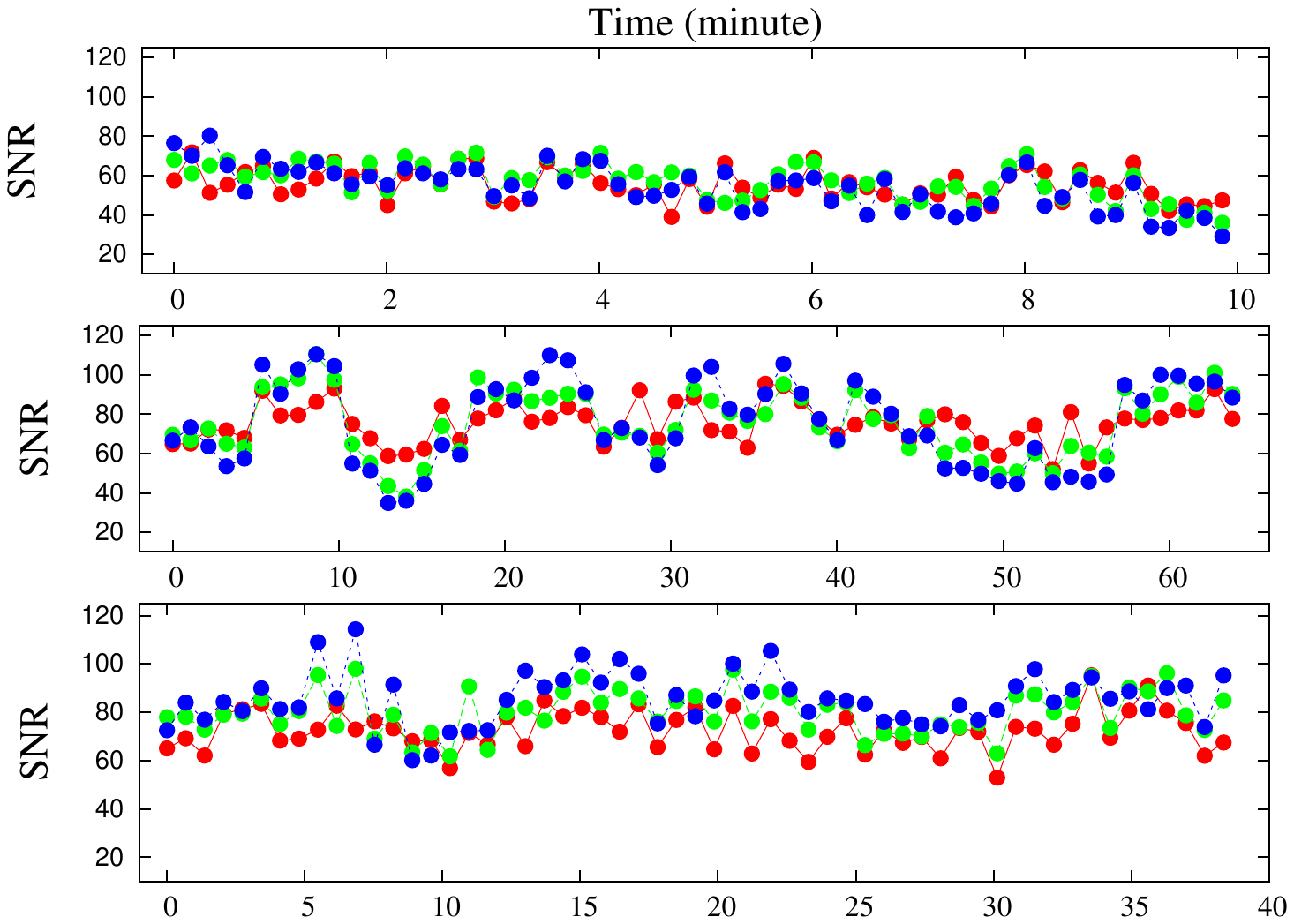}
 \caption[Sub-interval SNR variation for conventional and in-field phasing]
  {Sub-interval SNR variation over observation duration. Top panel showing
   SNR variation during observation with conventional phasing carried once in 
   beginning. Middle
   panel shows variation in sub-interval SNR during observation in which
   in-field phasing was carried regularly in background with 4.5 minutes 
   solution interval. Bottom panel shows variation in sub-interval SNR
   during observation with solution interval of 1.5 minutes. In all 
   panels, the color coding is the same as in Figure~\ref{fig:dephasing-phasing}
   }
  \label{fig:TST2615_30sep2022_J1120-3618_B3_snr_1int_VP_2}
\end{figure}
%----------------------------------------------------------
Figure~\ref{fig:TST2615_30sep2022_J1120-3618_B3_snr_1int_VP_2} shows the SNR variation with conventional and in-field phasing for a weak millisecond pulsar MSP J1120$-$3618 (which was discovered at the GMRT, \citep{Bhattacharyya22}). The top panel shows the variation of SNR with time for a conventionally phased observation. The middle panel shows the variation of SNR for in-field phasing with a time interval of 4.5 minutes between successive phase updates, while the third panel shows the variation of SNR for in-field phasing for a time interval of 1.5 minutes between successive phase updates. All of these observations carried in close succession. As before data is shown for three different beams, with the same colour coding as for Figure~\ref{fig:dephasing-phasing}. Please note that the time scale is different for the three different panels, with the top panel showing only 10 minutes of data. As can be seen, even over as short a time interval as 10 minutes, the SNR degrades quite significantly (to $\sim$50\% of its original value) in the case of conventional phasing. On the other hand, with in-field phasing the SNR does not show measurable degradation even over timescales $\sim$1 hour. For the observations with phase updates every 4.5 minutes (middle panel) one can see one quite prominent modulations of the SNR, although there is no clear long-term trend. Such modulations were also seen in some other test observations carried out with 4.5 minutes phase updates. We conjecture that this may be due to the ionosphere varying faster than the 4.5 minutes. Subsequently, we tried the phase update interval to 1.5 minutes, which as can be seen from the bottom panel of the figure results in a more constant SNR. 
%\textbf{ We have one more such observation, but with conventional phasing 
%and 4.5 minutes phasing interval for in-field phasing (i.e. without 1.5 
%minutes interval in-field phasing). However, all pulsar timing observations 
%are carried with $\sim$1.5 minutes solution interval and they exhibit flatter SNR.}
For all the observations presented later in the paper the update interval was kept at 1.5 minutes. 
\iffalse
%We estimated the scaled SNR for 10 minutes of all these parts in Fig. \ref{fig:TST2615_30sep2022_J1120-3618_B3_snr_1int_VP_2} and  we find $\sim$20\% SNR improvement for beam3 (blue curve, 24 antennas) from conventional phasing to in-field phasing with 1.5 minutes of solution interval, and $\sim$10\%  improvement in SNR for beam3 (24 antennas) with in-field phasing over beam2 (21 antennas) with conventional phasing. It is to be noted that beam2 was formed  with 21 antennas and beam3 was formed with 24 antennas and all arm antennas  are included in beam3.
\fi
%-------------------------------------------------------------
%\begin{figure}[!ht]\subsection{Science validation : application to MSPs}
\label{sec:Application2MSP}
\par
We present here results on the improvement that in-field phasing at the GMRT makes to two specific measurements of scientific interest, i.e. the accuracy with which pulsar timing can be done, as well as the difference that it makes to studies of eclipsing pulsars.

The stable clock-like properties of MSPs enables precise measurement of their rotational and orbital (if in binary) parameters through accurate measurement of the TOAs of pulses. For the same reasons, MSPs are also excellent probes of propagation imprints from interstellar medium (ISM) as well as detecting low-frequency gravitational waves (GWs) (e.g. \cite{Foster1990}).
For pulsar signal processing a mean stable pulse profile over the observing span is obtained by dedispersion and folding beam data with suitable model parameters (\cite{Lorimer2004}). The TOAs are determined by cross-correlation of the observed mean pulse profile of each epoch
with a high SNR template profile which is obtained by adding profiles from many observations performed at the same frequency (e.g. \cite{Taylor1992}).  The observed pulsar profile is a scaled and shifted version of template as described by equation \ref{eq:timing}
where $\tau$ signifies the shift in the measured profile, $a$ is the scaling factor and $N$ is the additive noise in the TOA measurement. The sensitivity of observation plays an important role in reducing the timing noise to obtain TOAs with improved uncertainties.   
\begin{equation}
\label{eq:timing}
P(t) = a + bT(t-\tau) + N(t)
%\label{eq:timing}
\end{equation}
%where $\tau$ signifies the ToA of measured profile from start of observation. 
%Sensitivity of observation plays important role and it introduces an uncertainty or
%contributes to timing noise in obtained ToA for less sensitive
%observations. 
The uncertainty in the TOA is given by the equation, $\sigma_{TOA} = \frac{W }{SNR}$, where $W$ is pulse width, and $SNR$ is sensitivity of profile. We present below in-field measurements of TOAs with the GMRT and compare the accuracy of the measurements with what is typically obtained via conventional phasing.
\par
The second measurement that we present is for an eclipsing millisecond pulsar. Millisecond binary pulsars are believed to be formed via recycling, a process in which the pulsar accretes mass from a companion star in binary, resulting in faster spin period via angular momentum transfer (see e.g. \cite{Bhattacharyya1992}). In a compact  binary system, the pulsar could also ablate mass from the companion via a  powerful pulsar wind. This  material can then obscure the pulsar emission causing eclipses
(e.g  \cite{Podsiadlowski1991}; \cite{VandenHeuvel1988}; \cite{Phinney1988}; \cite{Kluzniak1988}.
In general, the eclipse duration is dependent on the observation frequency and is typically longer at lower frequencies and shorter at higher frequencies.
 In many cases, the pulsar is detected throughout the entire orbit at higher frequencies where it shows eclipses at lower frequencies
 (e.g. \cite{Bhattacharyya13}).
 In order to understand eclipse mechanism, it is essential to precisely measure the frequency at which eclipse starts, i.e. the cut-off frequency. In addition, the pulse TOAs show an excess delay in pulse arrival time at the eclipse boundaries caused by the propagation of pulses through the eclipsing material spread beyond the Roche lobe (\cite{Eggleton1883}) of the companion star. This delay in TOAs arises from an excess $DM_{ex}$ associated with material that the radio pulses encounter at the eclipse boundaries as given by eqn.~\ref{eq:dm}, where $t_{ex}$ is delay in timing residual, $F$ is frequency of observation. In turn, $DM_{ex}$  can be used to compute electron column density $N_e$ of the eclipsing materials  (Eqn \ref{eq:ne}), as well as to constrain the eclipse mechanism.

  \begin{equation}
  \label{eq:dm} 
  %\mathrm{DM_{ex} ~ (pc~cm^{-3}) =  2.4 * 10 ^ {-10} * t_{ex} (\mu s) * F (MHz)^2}
  {DM_{ex} ~ (pc~cm^{-3}) =  2.4 * 10 ^ {-10} * t_{ex} (\mu s) * F (MHz)^2}
 \end {equation}
 
 \begin{equation}
 \label{eq:ne}
 {N_e (cm^{-2}) = 3 * 10^{18} * DM_{ex} (pc~cm^{-3})}
 \end{equation}
\par
We present here results for two binary pulsars MSP~J1120$-$3618 and  MSP 
J1544+4937. Both pulsars were discovered at the GMRT (Bhattacharyya et. al., 2022 and Bhattacharyya et. al., 2013 respectively) and have been the target of many follow up observations using conventional phasing with the uGMRT. This provides a good baseline for comparison between conventional and in-field phasing.
Table \ref{tab:observation_details_of_MSPs} lists the observation details for all the pulsars used for in-field phasing demonstrations.
\par
\begin{table}[ht!]
\caption{Observation details and parameters for MSPs}
\begin{tabular}{l r r r l r l}
\hline
Pulsar & Period   & Dispersion  & Flux     & Freq.  & No. of  & Obs.\\
       &          &  measure    & density  &            & epochs  & Span\\
      & mili-sec  & pc $cm^{-3}$       & mJy      &  MHz       &         & days\\ 
\hline
B0740$-$28   & 166.8 &  73.7 & 130   & 600.0 & 2  & NA  \\
J1120$-$3618 & 5.6 &  45.1 & 1    & 400.0 & 20 & 400 \\
J1544+4937 & 2.2 &  23.2    & 5.4   & 400.0 & 2  & NA  \\
\hline
\end{tabular}
\label{tab:observation_details_of_MSPs}
\end{table}

MSP J1120$-$3618 was discovered in the Fermi-directed survey with the GMRT and has a spin period of 5.6 ms, DM  of 45.1 pc cm$^{-3}$, binary period of 5.6 days, minimum companion mass of $~$0.18M$_{\odot}$, and flux density of $\sim$1 mJy at 400 MHz (\cite{Bhattacharyya22}). Decade long timing measurements for this system were presented by \cite {Sharma2024}  who detected a systematic increase of DM with time. We observed this MSP with in-field phasing at the GMRT band-3 (300-500 MHz) with 81.92 $\mu$s time resolution and 4096 spectral channels across 200 MHz bandwidth (i.e. 48.8 kHz frequency resolution). Both this data as well as the comparison conventional phasing data that we show were in-coherently dispersed. 

%
%---------------------------------------------------------------------
\subsubsection{MSP J1120$-$3618 : Improvement in timing precision}
%-------------------------------------------------------------
\begin{figure}[ht!]
\centering
\subfloat[]{\includegraphics[width=0.5\textwidth]{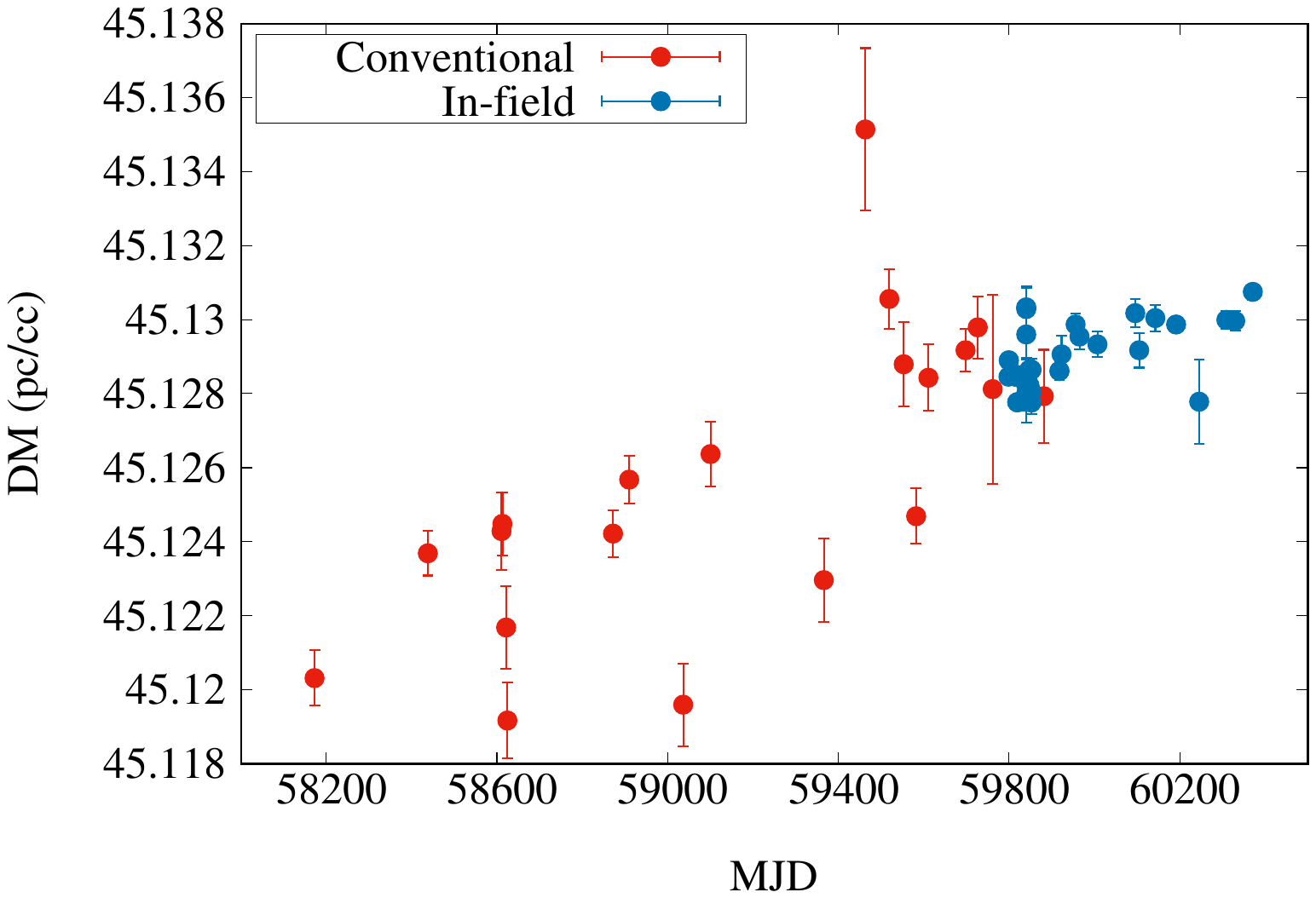}} 
\subfloat[]{\includegraphics[width=0.5\textwidth]{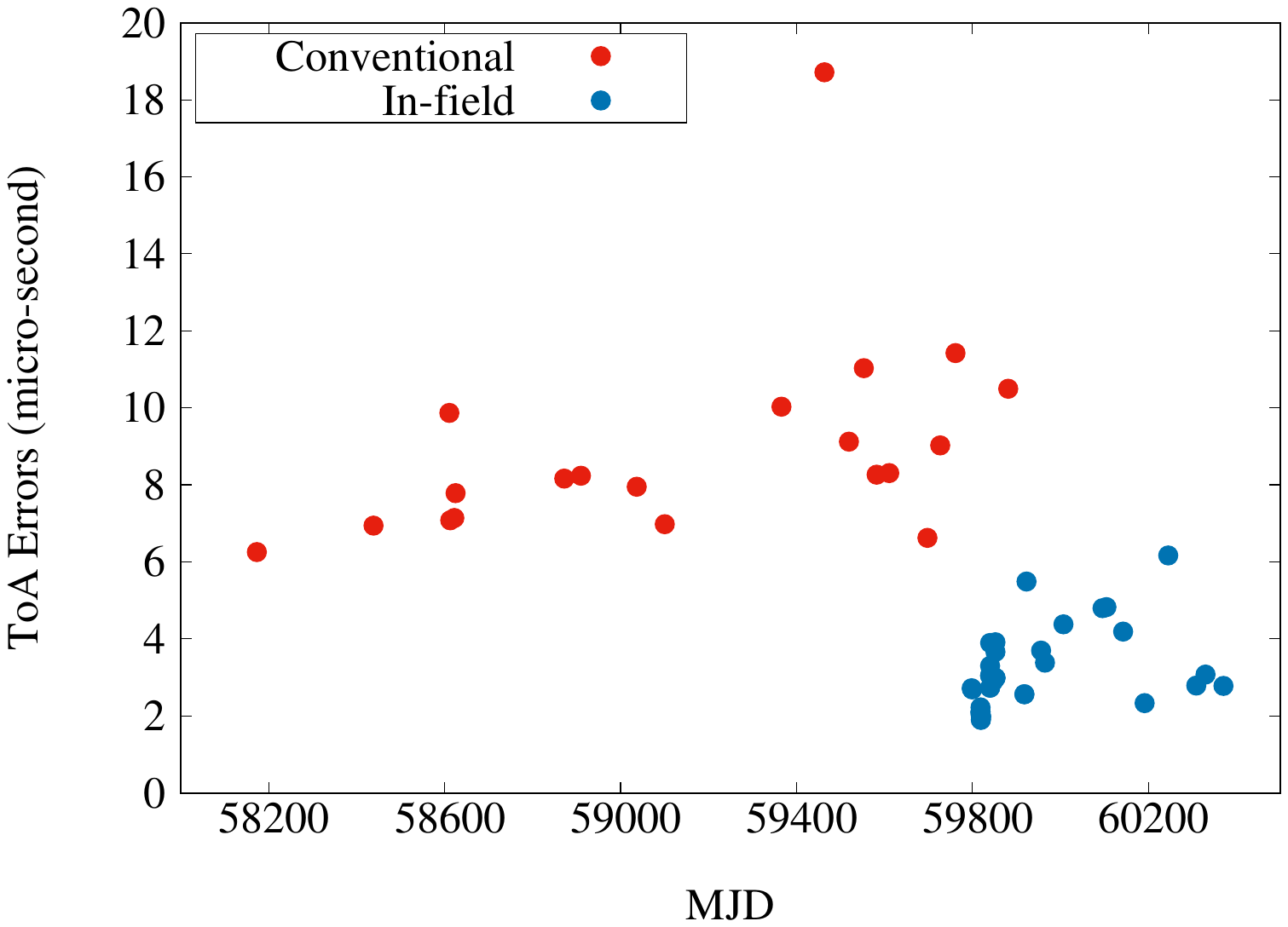}} 
 \caption{DM (Part (a)) and TOA errors (Part (b)) with conventional phasing  observations (red points) and in-field phasing observations (blue points) for MSP J1120$-$3618.}
\label{fig:dm_toa_error}
\end{figure}
%-------------------------------------------------------------
We show here continued timing of this pulsar, carried out over the last 
$\sim$1.6 years using in-field phasing. As discussed earlier, in-field phasing is expected to provide superior SNR not just because of improved phasing but also because the entire array is used in the observations, where as conventional phased arrays at the GMRT typically exclude the extreme arm antennas (i.e. using only about 70\% of the array). For ease of comparison all the data were scaled to a fiducial integration time of 30 minutes and a 100 MHz bandwidth. 
The Dispersion Measure was estimated from the time of arrivals obtained by dividing the 200 MHz bandwidth into 16 sub-bands and averaging full observation length
in one sub-integration. We carried PulsePortraiture-based  wide-band timing
analysis (\cite{Pennucci2014}, \cite{Sharma2022})  by modeling frequency-dependent 
effects and correcting for 
the evolution of the pulsar proﬁle with frequency as carried in
\cite{Sharma2024}.
We show in Fig. \ref{fig:dm_toa_error}(a)  the DM measurements obtained using in-field phasing (blue points);  we achieve a median DM precision of $\sim$2.9 x $10^{-4}$ pc cm$^{-3}$ which is about $\sim$3 times better than that obtained via conventional phasing (i.e. $\sim$8.9 x $10^{-4}$ pc cm$^{-3}$, red points). The TOA precision Fig. \ref{fig:dm_toa_error})(b) also shown improvement by a factor of $\sim$2.7. The median TOA error for the in-field phasing (blue points) is $\sim$3 $\mu$s, compared to $\sim$8.3 $\mu$s for conventional phasing (red points). The median SNR obtained from conventional phasing observations is $\sim$56 and median SNR from in-field phasing observations is $\sim$159 with their ratio of $\sim$2.8 close to the improvement we achieved in DM and TOA precision. 
%---------------------------------------------------------------------------
\subsubsection{MSP J1544+4937 : high time-frequency resolution probe for eclipse study}
\iffalse
The black-widow MSP J1544+4937 was discovered in a {\itFERMI}-directed search of unassociated gamma-ray sources.  It has a 2.16 ms spin period and is in a 2.9-hours of compact circular orbit with companion of mass 0.017M$_{\odot}$. Eclipses are detected at both 322~ MHz, and 607 ~MHz (Bhattacharyya, et. al., 2013). \cite{Kansabanik2021} carried a multi-frequency eclipse study and constrained frequency of onset of eclipse to be 345+/- 5 MHz. With the best-fit model to the eclipse phase spectra, they concluded that the eclipse mechanism is synchrotron
absorption. \cite{SKumari2023} carried out a decade-long timing study of this MSP and detected significant proper motion, temporal variation in DM, frequency-dependent DM, and revealed a secular variation of the orbital period. We observed this MSP using in-field phasing over 300-500~MHz, with 81.92$\mu$s time resolution and 4K channels over the 200MHz bandwidth.
%
\fi
%\par
We carried observations of two consecutive eclipses i.e., separated by one full orbit. One eclipse was observed with conventional phasing and the following one was with in-field phasing, both having $\sim$85 minutes of on-source time (spanning non-eclipsing orbital phases). We note that in the observations reported by \cite{SKumari2024}  two consecutive 
eclipses separated by an orbit have shown different characteristics, viz change
in eclipse cut-off frequency, excess $N_e$ etc. The SNRs (scaled to 1 hour integration) of the pulse profile at the non-eclipsing phase are $\sim$188 for the observation using conventional phasing and  
$\sim$265 for the consecutive observation with in-field phasing.
\par
\begin{figure}[!ht] 
\centering
      \includegraphics[width=3.2in,angle=0]{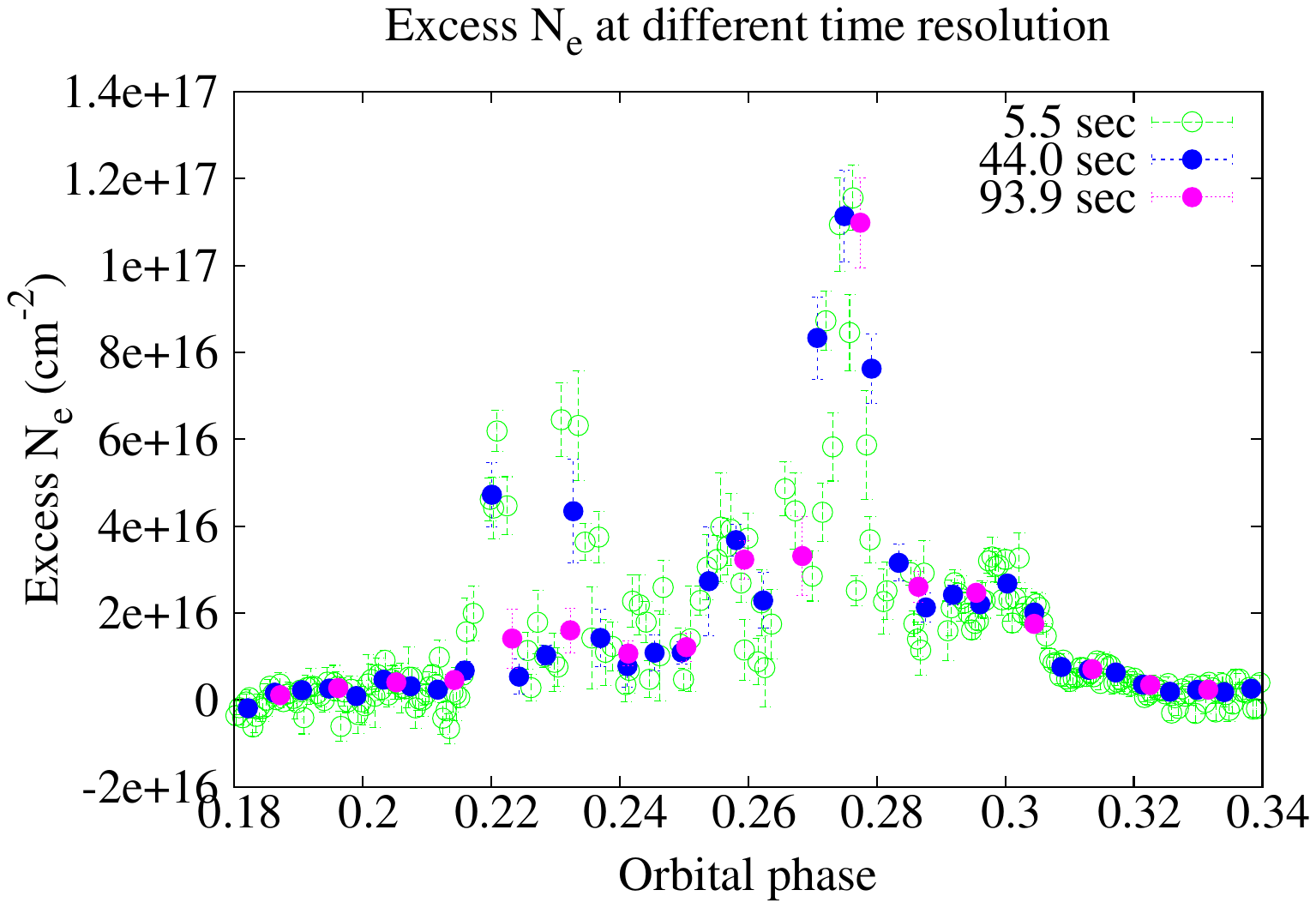}
	\caption[Measured excess electron column density for MSP J1544+4937 with
     in-field phasing for different time resolutions]
	{ Measured excess electron column density for MSP J1544+4937 with in-field
        phasing for different time resolutions.}
	\label{fig:J1544_13Jul2023_Ne}
\end{figure}

We performed another observation of J1544+4937 with in-field phasing where we could probe the eclipse region in much more detail. We recorded phased array data with all the available antennas (22) at the time of the observation. These included the outer arm antennas. Using the best-fit parameters obtained by \cite{SKumari2023}, we estimated the residuals across the eclipse (including some non-eclipsing orbital phases on either sides) region and derived the electron column density of the eclipsing materials N$_e$ by using equation 
\ref{eq:ne}.
%---------------------------------------
%\begin{math}
%$N_e$ = 3 $*$ $10^{18}$ $*$ 2.4 $*$ $10^{-10}$ $*$ ${R}$ $*$ ${F^2}$
%\label{eq:res2ne}
%\end{math}
%----------------------------------------
%where $t_{ex}$ ($\mu$s) is the best-fit model residual and $F$ is observation frequency (MHz). 
The excess electron density derived from the TOAs are shown in Fig. \ref{fig:J1544_13Jul2023_Ne}. \cite{SKumari2023} had shown similar excess TOAs  with a median time resolution of $\sim$60 seconds. Given our higher sensitivity we measured the excess N$_e$ at a range of time resolutions, viz. 93.9 seconds, 44.0 seconds and 5.5 seconds. As can be seen the higher time resolution observations show fine structure in the excess N$_e$, which is smoothed over at the coarser time resolution. 
This allows us to probe the variation of $N_e$ across the
eclipse and transitions in greater  detail exploring the possible clumpy nature of
eclipsing material around companion.
%and accurate measurement of the mass-loss rate of the companion can be possible using  $\dot{M}_c \sim \pi R^{2}_{E}m_{p}n_{e}V_{w}$,(\cite{Thompson1994}),  where $n_e = N_e/(2R_E)$, is the electron volume density.} 
%-------------------
\par
To obtain the eclipse cut-off frequency we followed the 
methodology described in \cite{RSharan2024}, in which data cube 
output of \textit{PRESTO} (i.e. pulse  intensity represented as function of  sub-integration, sub-bands and phase bins) is analysed for eclipse duration.  The finest frequency resolution (i.e. minimum number of sub-bands integration) of the data cube
are checked for which pulse is detected above 5-sigma significance. We measured 
the eclipse cut-off frequency  as 400 $\pm$ 1.6 MHz, which is a factor of $\sim 3$ times  more precise than the measurements reported by \cite{Kansabanik2021} and \cite{SKumari2024}). However, we note that the 7 MHz error bars on cut-off frequency used in \cite{SKumari2024} was based on the requirement of allowing uniform resolution for all eclipses spanning over a longer duration. The individual eclipse boosted by scintillation can have better frequency resolution.
%However we note that due to pulsar flux density variation caused by
%scintillation, the achievable time and frequency resolutions for eclipse study
%can be different for different epochs.
This increased precision can be translated to more stringent limits on the physical parameters of the eclipse medium. We plan in our future paper to describe the quantitative estimation of eclipse parameters aided with in-field phasing sampling an ensemble of eclipses. 
%--------------------------------------------------------------------------
%###############################################################################
\section{Summary}
\label{sec:Discussion}
%---------------------------------------
We present a scheme ("in-field phasing") for significantly improving the SNR of low frequency tied arrays where model visibilities of the target field are used to calibrate the array in quasi real time. This is expected to provide a number of improvements over conventional phasing where the array is periodically phased using observations of a calibrator near the target field.  Conventional phasing does not not correct for variations in the phase with time or position in the sky. Further, since the phase variation is typically most rapid on the distant baselines, for arrays like the GMRT, the most distant arm antennas are typically not used in tied arrays formed using conventional phasing. Finally since calibrator sources need to be periodically observed, conventional phasing does not allow for long contiguous observations of the target source. In-field phasing overcomes all of these drawbacks.

Using an implementation of in-field phasing at the upgraded GMRT (uGMRT) we show the improved performance of in-field phasing using observations of both normal as well as milli-second pulsars. For conventional phasing the signal to noise ratio (SNR) drops with time, with the SNR dropping further for arrays which include distant antennas, as expected.  In contrast the SNR remains constant with time for in-field phasing, even for arrays that include the most distant arm antennas. For the uGMRT the use of in-field observations is shown to significantly increase the signal to noise ratio, even for relatively short ($\sim$ 1hr) observations.
%--------------------------------------------------
\iffalse
\par
%We find from our observation of MSP J1120-3618 that array once phased in beginning by conventional method, showed drop of SNR over 10 minutes of observation to nearly half of its starting SNR (See figure \ref{fig:TST2615_30sep2022_J1120-3618_B3_snr_1int_VP_2}. top panel) Also, we find with phasing on a timescale of 4.5 minutes  interval, a systematic pattern consisting of rise and drop of  SNR of about 12–18 minutes cycle and changing the structure afterwards. During the rise of SNR, the largest beam showed the highest SNR among the three beams, and during the drop, the largest array  is affected most, showing the lowest SNR (see figure \ref{fig:TST2615_30sep2022_J1120-3618_B3_snr_1int_VP_2} middle panel. With a faster phasing  interval of $\sim$1.5 minutes, we obtained a higher SNR for  a larger array for a longer duration (as shown in figure  \ref{fig:TST2615_30sep2022_J1120-3618_B3_snr_1int_VP_2} bottom panel). For the largest array the SNR scaled to 10 minutes showed $\sim$20\% improvement for in-field phasing in comparison with conventional phasing.
\fi
\par
From observations of a milli-second pulsar (J1120$-$3618),  we show that the improved SNR in in-field phasing translates to an improved precision in the measurement of the dispersion measure (DM), viz.  a median DM precision of $\sim$2.9 x $ 10 ^{-4}~pc ~cm^{-3}$, which is $\sim$3 times better than the precision achieved using observations with conventional phasing.  The median scaled TOA precision  $\sim$3 $\mu$s, also shows an improvement in precision by a factor $\sim 3$. We also present observations of an eclipsing black-widow MSP J1544+4937 where in-field phasing allows for a more precise measurement of the eclipse cut-off frequency, as well as better measurements of the fine scale spatial structure of the eclipsing material as compared to earlier studies \cite{SKumari2024} and \cite{Kansabanik2021}.
\par
We acknowledge support of the Department of Atomic
Energy, Government of India, under project no. 12-R\&D-TFR-5.02-0700.
The GMRT is run by the National Centre for Radio
Astrophysics of the Tata Institute of Fundamental Research,
India. We acknowledge support of GMRT telescope operators
for observations. 
%---------------------------------------------------------------
%We acknowledge support of the Department of Atomic 
%Energy, Government of India, under project no. 12-R\&D-TFR-5.02-0700.
%The GMRT is run by the National Centre for Radio
%Astrophysics of the Tata Institute of Fundamental Research,
%India. We acknowledge support of GMRT telescope operators
%for observations.
%-------------------------------------------------------------------------------

%-------------------------------------------------------------------
\end{document}